\documentstyle[12pt]{article}
\input epsf
\newcounter{subequation}[equation]



\begin{document}
\title{\bf Equivariant self-similar wave maps from Minkowski spacetime
 into $3$-sphere }
\author{
  Piotr Bizo\'n\\
Institute of Physics,
Jagellonian University, Krak\'ow, Poland}

   \baselineskip=18pt
   \maketitle

\def\M{\Gamma^{-\alpha}}
\def\un{\underline}
\def\nn{\nonumber}
\newcommand{\no}{\nonumber}
\newcommand{\be}{\begin{equation}}
\newcommand{\ee}{\end{equation}}
\newcommand{\ba}{\begin{eqnarray}}
\newcommand{\ea}{\end{eqnarray}}
\newcommand{\ve}{\varepsilon}
\newcommand{\nit}\noindent
\newcommand{\D}\partial
\def\Pc{{\bf P_c}}
\def\a{\alpha}
\def\b{\beta}
\newcommand{\lan}\langle
\newcommand{\ran}\rangle
\newcommand{\wpl}{w_+}
\newcommand{\wmi}{w_-}
\newcommand{\vpsi}{\vec\psi}
\newcommand{\vg}{\vec g}
\newcommand{\la}{\lambda}
\newcommand{\ra}{\rightarrow}
\newcommand{\Z}{{\rm Z\kern-.35em Z}}
\newcommand{\bP}{{\rm I\kern-.15em P}}
\newcommand{\Q}{\kern.3em\rule{.07em}{.65em}\kern-.3em{\rm Q}}
\newcommand{\R}{{\rm I\kern-.15em R}}
\newcommand{\h}{{\rm I\kern-.15em H}}
\newcommand{\C}{\kern.3em\rule{.07em}{.55em}\kern-.3em{\rm C}}
\newcommand{\T}{{\rm T\kern-.35em T}}

\newtheorem{theo}{Theorem}
\newtheorem{defin}{Definition}
\newtheorem{prop}{Proposition}
\newtheorem{lem}{Lemma}
\newtheorem{cor}{Corollary}
\newtheorem{rmk}{Remark}

\begin{abstract}
\nit
We prove  existence of
 a countable family
of spherically symmetric self-similar wave maps from $3+1$
Minkowski
 spacetime into the 3-sphere. These maps can be
   viewed as excitations
 of the ground state wave map found previously by Shatah. The first
 excitation
is particularly interesting in the context of the Cauchy problem
 since it plays the role of a critical
 solution sitting at the threshold of singularity formation.
 We analyze the linear stability of our wave maps and show that
 the number of unstable modes about a given map is equal to its
 excitation index. Finally,
 we formulate a condition under which
  these results can be generalized to higher dimensions.
\end{abstract}
\thispagestyle{empty}

\section{Introduction.} Wave maps, defined as harmonic maps from a spacetime
$(M,\eta)$ into a Riemannian manifold $(N,g)$, have been intensively
studied during the past decade (see the recent review~\cite{struve}).
  The interest in wave
maps (sometimes called also sigma models) stems from the
fact that they contain many features of more complex relativistic
field models
 but are simple enough to be tractable rigorously.
 In particular, the investigation of  questions of global existence
and
 formation of
singularities for wave maps can give insight into the analogous,
but much more difficult, problems  in general relativity. With
this motivation we have recently studied numerically  the
development of singularities for wave maps from $3+1$ Minkowski
spacetime into the $3$-sphere~\cite{my}. In
 this case it was known that:$\;$ (i) solutions with  small
initial data exist globally in time~\cite{sid,kov};$\,\,$ (ii)
there exist smooth initial data which lead to blow-up in finite
time. An example of (ii) is due to Shatah~\cite{shatah} who
constructed a spherically symmetric self-similar wave map of the
form $u(r,t)=f_0(\frac{r}{T-t})$. This solution is perfectly
smooth for $t<T$ but it breaks down at $t=T$.
 Our numerical simulations~\cite{my} strongly suggest that the
 self-similar
 blow-up found by Shatah is
generic in the sense that there is a large set of initial data
which comprise the basin of attraction of  the solution $f_0$.
 In particular, it seems that all initial data of nonzero degree  (which by
definition are not small in the sense of~\cite{sid, kov}) blow up in
this universal self-similar manner.
The dynamical evolution of degree zero wave maps  is more interesting
because,
depending on the ``size'' of initial data, the solutions either exist
globally in time converging to the vacuum (this scenario is usually
referred to as dispersion),
 or blow-up in finite time
(where, as before, the blow-up profile is given by $f_0$). Thus,
in this case there arises a natural question of determining the
boundary between the basins of attraction of these two generic
asymptotic behaviors.
  In~\cite{my} we studied this question
numerically by evolving various one-parameter families of degree zero initial data
interpolating between blow-up and dispersion. A typical initial
data in this class is a gaussian with varying amplitude. We found that
the initial data lying on the boundary between the basins of
attraction of the solution $f_0$ and the vacuum solution converge
asymptotically to a certain codimension-one attractor which is self-similar.
This suggested that, besides $f_0$, the model admits  another self-similar
solution, call it $f_1(\frac{r}{T-t})$, which has exactly one unstable
 direction. This expectation was confirmed numerically in~\cite{my}.
In a sense, the  solution $f_1$ can be thought of as the
excitation of the ground state solution $f_0$.

The aim of this paper is to give a rigorous proof of existence of
a countable family of spherically symmetric self-similar wave maps
from Minkowski spacetime into the $3$-sphere. The above mentioned
solutions $f_0$ and $f_1$ are the first two elements of this
family. The proof is based on a shooting technique very similar to
the one used by us in the case of harmonic maps between
3-spheres~\cite{harm}.
\section{Preliminaries.}
A wave map $U$ is a map from a spacetime $M$ with metric $\eta$
into a Riemannian manifold $N$ with metric $g$ which is  a
critical point of the action
\be
L(U) = \frac{1}{2}\int_{M} g_{AB}\, \frac{\partial U^A}{\partial
x^a} \frac{\partial U^B}{\partial x^b} \, \eta^{ab} dV_M\, . \ee
The associated  Euler-Lagrange equations
\be
\Box_{\eta} U^A + \Gamma_{BC}^A(U) \partial_aU^B \partial^a U^C=0
\ee constitute the system of semilinear wave equations, where
$\Gamma$'s are the Christoffel symbols of the metric $g$. In this
paper we consider the case where $M=\R^{3+1}$, $3+1$ dimensional
Minkowski spacetime, and $N=S^3$, the unit $3$-sphere. In polar
coordinates on $\R^{3+1}$ and $S^3$ the respective metrics are
\be
\eta= -dt^2 + dr^2 + r^2 d\omega^2,
\ee
and
\be
g = du^2 + \sin^2(u) d\Omega^2,
\ee
where $d\omega^2$ and $d\Omega^2$ are the standard metrics on $S^2$,
and $u \in [0,\pi]$.
We consider spherically symmetric maps of the form
\be
U(t,r,\omega)= (u(t,r),\Omega=\omega). \ee Then the action (1)
reduces to
\be
L=\frac{1}{2} \int \left(-u_t^2+u_r^2+
  \frac{2\sin^2(u)}{r^2}\right) r^2 dt\, dr\, d\omega,
\ee
and the corresponding Euler-Lagrange equation is
\be
-u_{tt}+u_{rr}+\frac{2}{r} u_r - \frac{\sin(2u)}{r^2}=0.
\ee
This equation is invariant under dilations: if $u(t,r)$ is a solution
of equation (7), so is $u_{\lambda}(t,r)=u(\lambda t,\lambda r)$. It is thus natural
  to look for self-similar
solutions of the form
\be
u(t,r)=f\left(\frac{r}{T-t}\right),
\ee
where $T$ is a positive constant.
As mentioned in the introduction such solutions are important in the context
of the Cauchy problem for equation (7) since they
appear in the dynamical evolution as intermediate or final attractors.
Substituting the ansatz (8) into (7) we obtain the ordinary
differential equation
\be
f''+\frac{2}{\rho} f' -\frac{\sin(2f)}{\rho^2 (1-\rho^2)} = 0,
\ee
where $\rho=r/(T-t)$ and $'=d/d \rho$. For $t<T$
we have $0 \leq \rho< \infty$.

It is sufficient to consider equation (9) only inside the past
light cone of the point $(T,0)$, {\em i. e.}, for $\rho
\in [0,1]$. This constitutes the two-point singular boundary value problem with
the boundary conditions
\be
f(0)=0 \quad
 and \quad f(1)=\frac{\pi}{2},
\ee
which are dictated by the requirement of smoothness at the endpoints.
Once a solution of equation (9) satisfying the conditions (10) is
constructed, it
 can be
easily extended to $\rho>1$~\cite{shatah}. Note that solutions of
(9) and (10) are the critical points of the functional
\be
E[f] = \frac{1}{2} \int_0^1 \left( \rho^2 f'^2 -
  \frac{2 \cos^2(f)}{1-\rho^2} \right) d\rho,
\ee which, as was pointed out by Shatah and
Tahvildar-Zadeh~\cite{tahv}, can be interpreted as the energy for
harmonic maps from the hyperbolic space $H^3$ into the upper
hemisphere of $S^3$. Shatah~\cite{shatah} showed that $E[f]$ is
bounded from below over the $H^1$-space of functions satisfying
(10) and attains an infimum
 at a smooth function  $f_0$, the
 ground state solution of equation (9). Independently,
 Turok and Spergel~\cite{turok} found this solution in  closed form
\be
f_0=2 \arctan(\rho).
\ee

The central  result of this paper is
 \begin{theo}
There exist a countable family of smooth solutions $f_n$ of equation (9)
satisfying
the boundary conditions (10).
The index $n=0,1,2,...$ denotes the number of intersections of $f_n(\rho)$
with the line
$f=\pi/2$ (the equator of $S^3$) on $\rho\in[0,1)$.
\end{theo}
Before proving this theorem in the next section, we present now  some
numerical results.
 As will be shown below the solutions satisfying $f(1)=\pi/2$ form a
 one-parameter family with asymptotics $f(\rho)\sim \pi/2+ b (\rho-1)$
  near $x=1$,
while the solutions satisfying $f(0)=0$ form a one-parameter
family with asymptotics $f(\rho)\sim a \rho$ near $\rho=0$. The
solutions $f_n$ are obtained by a standard
shooting-to-a-fitting-point method, that is by integrating
equation (9) away from the singular points $\rho=0$ and $\rho=1$
in the opposite directions with some trial parameters $a$ and $b$
and then adjusting these parameters so that the solution joins
smoothly at the fitting point. The discrete set of pairs
$(a_n,b_n)$ generated in this way and the energies characterizing
the solutions $f_n$ are shown below for $n\leq 4$. \vskip 0.2cm $$
\begin{tabular}{|c|c|c|c|c|} \hline
$n$ & $a_n$ & $b_n$ & $E_n=E[f_n]$ & $E_n/E_{n+1}$ \\
 \hline
0 & 2 & 1 & $\pi/4-1$ & 10.891  \\
1 & 21.757413 & -0.305664 & -1.97045  $\times 10^{-2}$ & 10.764\\
2 & 234.50147 & 0.0932163  & -1.83055  $\times 10^{-3}$ & 10.751\\
3 & 2522.0683 &-0.0284312 & -1.70276  $\times 10^{-4}$ & 10.749\\
4 & 27113.388 & 0.0086717 & -1.58411 $\times 10^{-5}$ & 10.749 \\
 \hline
\end{tabular}
$$
\begin{figure}
\epsfxsize=11cm \centerline{\epsfbox{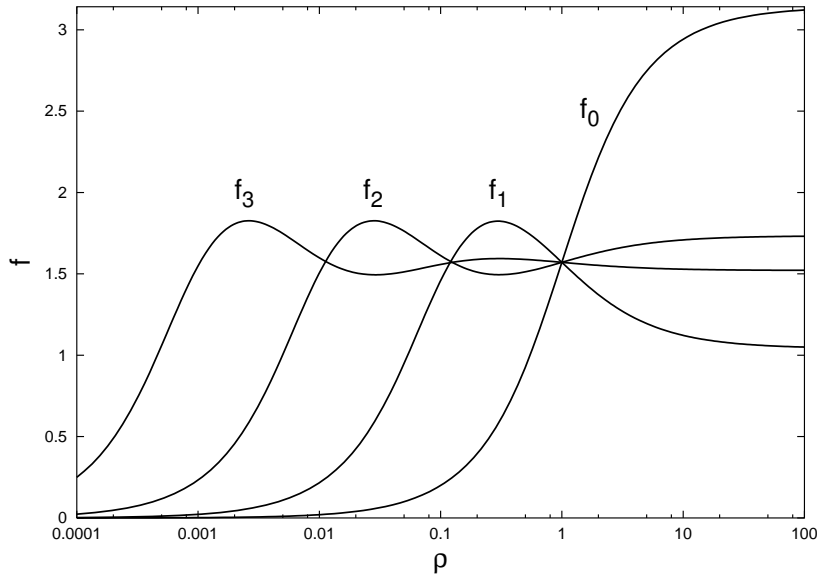}} \caption{The
ground state solution $f_0$ and the first three
  excitations generated numerically. The solutions $f_n$ with
  $n>0$ were first discovered numerically by \"Aminneborg and
  Bergstr\"om~\cite{ab}.}
  \end{figure}
\section{Proof of Theorem~1}
To prepare the ground for the proof of Theorem~1 we first discuss
some basic properties of solutions of equation (9). It is
convenient  to use  new variables defined by
\be
\rho=\frac{1}{\cosh{x}}, \quad {\mbox and } \quad
  h(x)=f(\rho)-\frac{\pi}{2}.
\ee
The range of $x$ is from $x=0$ (for $\rho=1$) to $x=\infty$ (for
$\rho=0$). Note that the number of intersections of $f$ with the line
$f=\pi/2$ is  the same as the number of zeros of $h$. In these new
variables equation (9) becomes
\be
h''-\coth(x) h' +\sin(2h)=0,
\ee
and the boundary conditions (10) translate into
\be
h(0)=0 \quad {\mbox and} \quad h(\infty)=\pm \frac{\pi}{2},
\ee
where the $\pm$ sign in the last expression, obviously allowed by the
reflection symmetry $h\ra-h$, is introduced for convenience.
\begin{lem} For any $b$ there exists a unique global solution $h_b(x)$ to
  equation (14)
  such that
\be
h_b(x)\sim b x^2
\ee
 as $x\ra 0$.
\end{lem}
{\em Proof:} Defining $v=h'$, let us rewrite equation (14) as the system of two
integral equations
\be
v(x)=-\sinh(x) \int_0^x \frac{\sin(2h(s))}{\sinh{s}} ds, \quad
h(x)=\int_0^x v(s) ds.
\ee
Following the standard procedure we solve (17) by iteration, setting
\be
v^{(n+1)}(x)=-\sinh(x) \int_0^x \frac{\sin(2h^{(n)}(s))}{\sinh{s}} ds, \quad
h^{(n+1)}(x)=\int_0^x v^{(n)}(s) ds.
\ee
with the starting values $h^{(0)}=bx^2$ and $v^{(0)}=2bx$.
 It can easily be shown that the mapping
$(h^{(n)}(x),v^{(n)}(x)) \ra (h^{(n+1)},v^{(n+1)}(x))$ defined by
(18) is contractive for any finite $x$, hence the sequence
$(h^{(n)},v^{(n)})$ converges to a solution of equation (14). The
proof of uniqueness is also routine so we omit it.
\begin{defin}
A solution of equation (14) starting at $x=0$ with the
asymptotic behavior (16) will be called the {\bf $b$-orbit}. Without loss of
 generality we assume that $b\geq 0$. The $b$-orbit which satisfies
$h(\infty)=\pm \pi/2$ will be called a {\bf connecting orbit}.
\end{defin}
{\em Remark 1:} In the following whenever we say ``a solution'' we
always mean the
$b$-orbit. Also, when we say that some property holds for all $x$ we
always mean for all $x>0$. We use
$\lim$ to denote $\lim_{x\ra\infty}$.\\
{\em Remark 2:} The endpoints of connecting orbits $(h=\pm \pi/2,h'=0)$
are saddle-type critical points of the asymptotic ($x\ra\infty$) autonomous equation
$h''-h'+\sin{2h}=0$. One can easily show ({\em cf.}~\cite{harm}) that
the connecting orbits converge to these points along the
one-dimensional stable manifolds $\pm h(x) \sim -\pi/2+a e^{-x}$.

The following function, defined for $b$-orbits, will play a crucial
 role in our analysis
\be
W(x) = \frac{1}{2} h'^2 +\sin^2\!{h}.
\ee
We have
\be
\frac{dW}{dx}= \coth(x) h'^2,
\ee
so $W$ is increasing (unless $h$ is a constant solution).
Equations (19) and (20) imply that if $W(x_0)\geq 1$ for some $x_0$ (
and $h$ is not identically equal to $\pm \pi/2$) then
$|h'(x)|>\epsilon>0$ for $x>x_0$, hence $\lim W(x)=\infty$.
 Thus, if a $b$-orbit crosses the line $h=\pm \pi/2$,
then $h'$ and $h$ tend monotonically to $\pm\infty$.
\begin{lem}
A $b$-orbit (with nonzero $b$) which satisfies  $|h(x)| <\pi/2$ for all
 $x$, is a connecting orbit.
\end{lem}
\nit {\em Proof:} We showed above that if  $W(x_0)\geq 1$ for some
$x_0$, then $|h|$ tends to infinity, hence
$|h|<\pi/2$ implies that $W(x)<1$ for
all $x$, so $\lim W(x)$ exists. Thus, $\lim W'=0$
which means by (20) that $\lim h'=0$ and next by (19) that $\lim
\sin^2\!{h}$ exists, implying
that also  $\lim h$ exists. By
equation (14), $\sin{2h(\infty)}=0$ since otherwise $\lim h''\neq 0$
contradicting $\lim h'=0$. Hence, $h(\infty)=\pm \pi/2$ or $
h(\infty)=0$. To conclude the proof note that
the latter implies $\lim W=0$ which in view of (20) is possible only if $W \equiv 0$, that is  $h \equiv 0$.

The next two lemmas describe the behavior of $b$-orbits for small
and large values of the shooting parameter $b$, respectively.
\begin{lem} If $b$ is sufficiently small then
 the solution $h_b(x)$ has arbitrarily many zeros.
\end{lem}
\nit {\em Proof:} Define $\tilde h(x)=h_b(x)/b$. The function
$\tilde h$ satisfies
\be
\tilde h'' -\coth(x) \tilde h' +\frac{\sin(2 b\tilde h)}{b}=0 \ee
with the asymptotic behavior $\tilde h(x) \sim x^2$ as $x \ra 0$.
As $b \rightarrow 0$, the solutions of equation (21) tend
uniformly on compact intervals to the solution of the limiting
equation
\be
H'' - \coth(x) H' +2 H=0
\ee
with the asymptotic behavior $H(x)\sim x^2$ as $x\ra 0$. The solution
$H(x)$
can be found in closed form in terms of the hypergeometric function but for the purpose of the argument it
is enough to observe
that $H(x)$ is oscillating at infinity, since this implies  that the
number of zeros of
 $h_b(x)=b\tilde
h(x)$ increases to infinity as $b$ tends to zero.
\begin{lem}
If $b$ is sufficiently large then the solution $h_b(x)$ increases
monotonically to $\infty$.
\end{lem}
\nit {\em Proof:} As in the proof of Lemma~3, we use a scaling
argument. This time, we define $\bar h(x)=h_b(x/\sqrt{b})$. The
function $\bar h$ satisfies
\be
\bar h'' -\frac{1}{\sqrt{b}} \coth\left(\frac{x}{\sqrt{b}}\right)
\bar h' +\frac{\sin(2 \bar h)}{b}=0 \ee with the asymptotic
behavior $\bar h(x) \sim x^2$ as $x \ra 0$. As $b \ra \infty$, the
solutions of equation (23) tend uniformly on compact intervals to
the solution of the limiting equation
\be
\bar H'' - \frac{1}{x}\bar H'=0, \ee that is to $\bar H(x)=x^2$.
Thus, on any compact interval the solution $h_b(x)$ stays
arbitrarily close to $b x^2$ if $b$ is sufficiently large. In
particular, $h_b(x)$ strictly increases up to some $x_0$ where
$h(x_0)=\pi/2$. Since $W(x_0)>1$, by the argument following (20)
$h_b$ tends monotonically to $\infty$.
\vskip 0.2cm Now we are ready to prove Theorem~1. The proof will
the immediate corollary of the following proposition
\begin{prop}
There exists a decreasing sequence of positive numbers $\{b_n\}$, $n=0,1,2,...$,
such that the corresponding $b_n$-orbits are connecting orbits with
exactly $n$ zeros for $x>0$. Morever, $\lim_{n\ra\infty}b_n=0$.
\end{prop}
{\em Proof:} The proof is based on an inductive application of the
standard shooting argument.  Let $S_0=\{b\,|\,$ $h_b$ strictly increases up
 to some
  $x_0$ where $h_b(x_0)=\pi/2\}$. Let $b_0=\inf S_0$. By
Lemma~4 the set $S_0$ is nonempty and by Lemma~3 $b_0>\epsilon>0$.
The $b_0$-orbit
cannot cross the line $h=\pi/2$ at a finite $x$ because the same
would be true for nearby $b$-orbits with $b<b_0$, violating the
definition of $b_0$.  Thus, the $b_0$-orbit stays in the region
$|h|<\pi/2$ for all $x$, and therefore due to Lemma~2 it is a connecting orbit.
By definition the $b_0$-orbit has no zeros for $x>0$.

To make the inductive step we need one more lemma.
\begin{lem}
If $b=b_0-\epsilon$ for sufficiently small $\epsilon>0$, then the
solution $h_b(x)$ increases up to some $x_0$ where it attains a
positive local maximum $h(x_0)<\pi/2$ and then decreases monotonically
to $-\infty$.
\end{lem}
\nit {\em Proof:} By the definition of $b_0$ there must exist a point
 $x_0$ where $h_b'(x_0)=0$. Since by (14) a solution $h$ cannot have
 a local minimum if $h>0$, it follows that there must be
a point $x_1>x_0$ where $h_b(x_1)=0$ (otherwise the $b$-orbit would
 contradict Lemma~2). The idea of
the proof is to show that $W(x_1)>1$ provided that $\epsilon$ is
sufficiently small. As argued above this implies that for $x>x_1$
$h_b$ decreases monotonically to $-\infty$. In the following we
drop the index $b$ on $h_b$. From (19) we have
\be
W(x_1)-W(x_0)=\int_{x_0}^{x_1} \coth(x) h'^2 dx > -\int_0^{h(x_0)} h'
dh.
\ee
In order to estimate the last integral note that for $x>x_0$
\be
W(x)-W(x_0)=\frac{1}{2}h'^2+\sin^2\!{h(x)}-\sin^2\!{h(x_0)} > 0,
\ee
so $-h'>\sqrt{2(\sin^2\!{h(x_0)}-\sin^2\!{h})}$. Inserting this into
 (24) gives
\be
W(x_1)>\frac{1}{2} \sin^2\!{h(x_0)} + \int_0^{h(x_0)} \sqrt{
2(\sin^2\!{h(x_0)}-\sin^2\!{h})}\, dh.
\ee
The right-hand side of this inequality is an increasing function of
$h(x_0)$ which  exceeds $1$ if
$\pi/3<h(x_0)<\pi/2$, as can be checked by direct calculation.
The value $h_b(x_0)$ will fall into that
 interval if $\epsilon$ is sufficiently small because by
continuous dependence of solutions on initial conditions,
$h_b(x_0) \ra \pi/2$ as $\epsilon\ra 0$.
This concludes the proof of Lemma~5.
\vskip 0.2cm
Having Lemma~5 we return now to the proof of Proposition~1. Let
$S_1=\{b\,|\,
  h_b(x)$ increases up to some $x_0$ where it attains a positive local
  maximum $h(x_0)<\pi/2$ and then decreases monotonically up to some
  $x_1$ where $h(x_1)=-\pi/2$\}. Let $b_1=\inf S_1$. Due to Lemma~5
 the set $S_1$ is nonempty and by Lemma~3 $b_1$ is strictly positive.
Using the same
 argument
as above we conclude that the $b_1$-orbit must stay in the region
 $|h|<\pi/2$ for all $x$, so it is a connecting orbit
(asymptoting to $-\pi/2$). By definition
 the $b_1$-orbit has exactly one zero for $x>0$.

The subsequent connecting orbits are obtained by the repetition of
the above construction. Since the sequence $\{b_n\}$ is decreasing
and bounded below by zero, it has a nonnegative limit. Suppose
that $\lim_{n\ra\infty}b_n=b^*>0$. The $ b^*$-orbit cannot leave
the region $|h|<\pi/2$ for a finite $x$ because the set of such
orbits is clearly open. Thus, the $b^*$-orbit is a connecting
orbit with some finite number of zeros. But this contradicts the
fact that the number of zeros of $b_n$-orbits increases with $n$.
We conclude therefore that $\lim_{n\ra\infty} b_n = 0$. This
completes the proof of Proposition~1.
 Returning to the
original variables $f(\rho)$ and $\rho$, and using the notation
 $h_n(x) \equiv h_{b_n}(x)$, we have
$f_n(\rho)=h_n(x)+\pi/2$
with $f_n(1)=\pi/2$ and $f_n(0)=0 (mod\, \pi)$, as claimed in Theorem~1.

We end this section with a remark  about the large $n$ limit. From
 $\lim_{n\ra\infty}b_n=0$, it follows
that $\lim_{n\ra\infty}h_n(x)= 0$ for any finite $x$. The limiting
solution $h^*=0$ (or $f^*=\pi/2$) is a singular map which
geometrically corresponds to the map into the equator of $S^3$.
The ``energy'' of this map $E[f^*]=0$ provides the upper bound for
the ``energies'' of critical points of (11) (we write ``energy''
is quotation marks to emphasize that the functional (11) is not
the true conserved energy  associated with the action (6)). As
follows from the proof of Lemma~3, the behavior of connecting
orbits with large $n$ (and consequently small $b_n$) can be
approximated by the solution of equation (22), namely $h_n(x)
\approx b_n H(x)$ on $x\in[0,x_n)$ where  $x_n$ tends to infinity
as $n\ra\infty$. This fact can be used to prove some  remarkable
scaling properties of connecting orbits in the limit of large $n$.

For example one can show that (see the table in Section~2)
\be
\lim_{n\ra\infty} \frac{E_n}{E_{n+1}} =e^{\frac{2\pi}{\sqrt{7}}}.
\ee For more detailed discussion of this issue we refer the reader
to~\cite{harm} where the analogous behavior in the case of
harmonic maps between spheres was derived.
\section{Stability}
The role of self-similar solutions $f_n$ in the evolution depends
crucially on their stability with respect to small perturbations.
This problem was analysed by us in~\cite{my} by mixed
analytic-numerical methods. In particular, we provided evidence
towards the conjecture that the solution $f_0$ is asymptotically
stable and, as such, has an open basin of attraction. To make the
results obtained in~\cite{my}  rigorous is a formidable task. In
this section we discuss the first (easy) step in achieving this
goal, namely we determine the character of the spectrum of the
linearized operator around the solutions $f_n$. A somewhat
different but equivalent version of the linear stability analysis
was presented in~\cite{my}.

 We restrict
attention to the interior of the past light cone of the
 point $(T,0)$ and define the new time coordinate
 $s=-\ln{\sqrt{(T-t)^2-r^2}}$.
In terms of $s$ and $\rho$, Eq.(7) becomes
\begin{equation}
-\frac{e^{2 s}}{(1-\rho^2)^2}  (e^{-2 s} u_s)_s + u_{\rho\rho}
+\frac{2}{\rho} u_{\rho} -\frac{ \sin(2 u)}{\rho^2 (1-\rho^2)} =0.
\end{equation}
Of course, this equation reduces to Eq.(7) if the solution is
self-similar, that is $s$-independent. Following the standard
procedure we seek solutions of (29) in the form
$u(s,\rho)=f_n(\rho)+ w(s,\rho)$. Neglecting the $O(w^2)$ terms we
obtain a linear evolution equation for the perturbation
$w(s,\rho)$
\begin{equation}
-\frac{ e^{2 s}}{(1-\rho^2)^2} (e^{-2 s} w_s)_s + w_{\rho\rho}
+\frac{2}{\rho} w_{\rho} -\frac{ 2 \cos(2 f_n)}{\rho^2
(1-\rho^2)}\: w = 0.
\end{equation}
Substituting $w(s,\rho)=e^{(\lambda+1) s} v(\rho)$ into (30) we
get the eigenvalue problem
\begin{equation}
A v = (1-\lambda^2) v, \quad \mbox{where} \quad
A=-\frac{(1-\rho^2)^2}{\rho^2} \frac{d}{d\rho}\left(\rho^2
\frac{d}{d\rho}\right) + \frac{2 (1-\rho^2)\cos(2f_n)}{\rho^2}.
\end{equation}
Note that the principal part of the operator $A$ is the radial
Laplacian on the hyperbolic space  $H^3$.
 We consider this problem in the space of
functions which are square-integrable on the interval $[0,1]$ with
respect to the natural inner product on $H^3$, that is
\be
v \in L^2([0,1],\frac{\rho^2}{(1-\rho^2)^2} d\rho). \ee In this
function space $A$ is self-adjoint hence its spectrum is real.
Both endpoints are of the limit-point type.
 Near
$\rho=0$ the leading behavior of solutions of (31) is $v(\rho)
\sim \rho^{\alpha}$ where $\alpha(\alpha+1)=2$, so admissible
solutions  behave as
\begin{equation}
 v(\rho) \sim \rho \quad \mbox{as} \quad \rho\rightarrow 0.
 \end{equation}
Near $\rho=1$ the leading behavior is $v(\rho) \sim
(1-\rho)^\beta$ where $\beta=(1\pm\sqrt{\lambda^2})/2$ so
eigenfunctions must have $\lambda^2>0$ and behave as (up to a
normalization constant)
\begin{equation}
  v(\rho) \sim
   (1-\rho)^{\frac{1}{2} (1+|\lambda|)} \quad \mbox
  {as} \quad \rho\rightarrow 1.
\end{equation}
All $\lambda^2\leq 0$ belong to the continuous spectrum of $A$.
The case $\lambda=0$ will be treated separately below. Note that
this eigenvalue problem has the symmetry $\lambda \rightarrow
-\lambda$ (that is why we wrote $\lambda+1$ rather than $\lambda$
in the ansatz for $w$). Each eigenvalue $\lambda^2>0$ gives rise
to an unstable mode which grows exponentially as
$e^{(|\lambda|+1)s}$. To find the eigenvalues we need to solve
(31) on the interval $\rho\in [0,1]$ with the boundary conditions
(33) and (34). In~\cite{my} we did this numerically (for $n\leq
4$) by shooting the solutions from both ends and matching the
logarithmic derivatives at a midpoint. For example, for $n=1$ we
got $\lambda^2_1\approx 28.448$; for $n=2$ we got $\lambda^2_1
\approx 28.132, \lambda^2_2\approx 3372.12$. Our numerics strongly
suggested that the point spectrum of the operator $A$ around the
solution $f_n$ has exactly $n$ positive eigenvalues
$\lambda_k^2>0$ ($k=1,\ldots,n$). Now, we will give a simple proof
of this property.

The proof is based on the observation that the solution with
$\lambda=0$ corresponds to the gauge mode
 which is due to the freedom
 of choosing the blowup time $T$. To see this, consider a
solution $f_n(r/(T'-t))$. In terms of the similarity variables
$s=-\ln{\sqrt{(T-t)^2-r^2}}$ and $\rho=r/(T-t)$, we have
\begin{equation}
f_n\!\left(\frac{r}{T'-t}\right)=f_n\!\left(\frac{\rho}{1+\epsilon
\sqrt{1-\rho^2}\: e^s}\right) \quad \mbox{where} \quad
\epsilon=T'-T.
\end{equation} In other words, each self-similar solution $f_n(\rho)$
generates the orbit of solutions of (29) parametrized by
$\epsilon$.
 It is easy to verify that the generator of this orbit
\begin{equation}
  w(s,\rho) = -\frac{d}{d\epsilon}\:
  f_n\!\left(\frac{\rho}{1+\epsilon \sqrt{1-\rho^2}\:
  e^s}\right)\Big|_{
  \epsilon=0} = e^s \rho \sqrt{1-\rho^2} f'_n(\rho)
\end{equation}
solves (30), thus $v_{gauge}^{(n)}(\rho)=\rho \sqrt{1-\rho^2}
f'_n(\rho)$ satisfies (31) with $\lambda=0$\footnote{We emphasize
that $\lambda=0$ is not an eigenvalue because
$v_{gauge}^{(n)}(\rho)$ is not square-integrable at $\rho=1$.
Hovewer, this solution is distinguished from the rest of the
continuous spectrum by the fact that it is subdominant at $\rho=1$
(such a solution is sometimes referred to as a
\emph{pseudo-eigenfunction}).}. Since $v_{gauge}^{(n)}(\rho)$ has
exactly $n$ zeros on $\rho\in (0,1)$ (because $f_n$ has $n$
extrema), it follows by the standard result from Sturm-Liouville
theory that there are exactly $n$ positive eigenvalues, as
conjectured in~\cite{my}. To summarize, we showed that the
self-similar solution $f_n$ has exactly $n$ unstable modes, which
means in particular that the solution $f_0$ is linearly stable.
\vskip 0.2cm
\noindent{\em Remark.} If we view the solutions $f_n$ as harmonic
maps from the hyperboloid $H^3$ into $S^3$, then the eigenvalue
problem (31) determines the spectrum of the  Hessian of the energy
functional (11)
\be
\delta^2E[f_n](v,v) = \frac{1}{2} \int_0^1 \left( \rho^2 v'^2 +
  \frac{2 \cos(2 f_n)}{1-\rho^2}\: v^2 \right) d\rho.
\ee Within this approach the argument given above can be rephrased
in terms of the Morse index. In particular, it implies that the
Morse index of the solution $f_0$ is zero, in agreement with
Shatah's result that $f_0$ is a local minimum of the energy
functional (11). Note that in this context
 the gauge mode acquires  a geometrical
interpretation as the perturbation induced by $K$, the conformal
Killing vector field on $H^3$,
\be
v_{gauge}^{(n)}=\pounds_K f_n, \quad \mbox{where} \quad K=\rho
\sqrt{1-\rho^2}\: \partial/\partial\rho\;. \ee Aside, we remark
that by solving (31) one can show that the Morse index of the
singular map $f^*=\pi/2$ is infinite. This fact could  be probably
used to give an alternative proof of Theorem~1 via Morse theory
methods using the ideas of Corlette and Wald developed recently in
the case of harmonic maps between spheres~\cite{cw}.
\section{Generalization to higher dimensions}
The proof of Theorem~1 is rather robust which suggests that the
result can be  generalized in various directions. One possibility,
which will not be pursued here, is to consider more general
nonconvex targets\footnote{For example, one can easily verify that
the proof of Theorem~1 goes through if the metric (4) is replaced
 by $g=du^2 + s^2(u)d\Omega^2$ where the function $s(u)$
  satisfies the following conditions (cf.~\cite{csz}): (i)
 $s(0)=0$ and $s'(0)=1$; (ii) $s(u)$ is monotone increasing from $u=0$ up
 to some $u^*>0$ where it attains a maximum.}.
Another possibility is to consider the analogous problem in higher
dimensions, that is wave maps $U: M \rightarrow N$, where
$M=\R^{m+1}$, $m+1$ dimensional Minkowski spacetime, and $N=S^m$,
the unit $m$-sphere. At the same time one can relax the
equivariance ansatz (5) by admitting the maps of the form
\be
U(t,r,\omega)= (u(t,r),\Omega=\chi(\omega)), \ee where $\chi$ is a
homogeneous harmonic polynomial of degree $l>0$. The ansatz (5) is
the special case of (39) with $l=1$.
 Assuming
self-similarity we obtain the analogue of Eq.(9)
\be
f''+\left(\frac{m-1}{\rho}+\frac{(m-3)\rho}{1-\rho^2}\right) f'
-\frac{k \sin(2f)}{\rho^2 (1-\rho^2)} = 0,
 \ee
where $k=l(l+m-2)/2 $. As before, we want to construct smooth
solutions on the interval $[0,1]$ satisfying the boundary
conditions (10). Standard analysis of the behavior of such
solutions at the endpoints yields that $f(\rho) \sim a \rho^l$
near $\rho=0$ and $f(\rho) \sim \pi/2+ b (1-\rho)^{(m-1)/2}$. Note
that the latter implies that the desired smooth solutions can
exist only if the dimension $m$ is odd. Although Eq.(40) looks
more complex than (9), the same change of variables as in (13)
transforms (40) into
\be
h''-(m-2) \coth(x) h' + k \sin(2h)=0, \ee which has the same form
as (14) apart from the change of constant coefficients. Now, let
us see which steps of the shooting argument from Section~3 are
effected by this change of coefficients. Lemma~1 holds with the
asymptotic behavior near $x=0$ replaced by $h(x)\sim b x^{m-1}$.
Lemmas~2 and 4 remain valid because their proofs are dimension
independent. The only fact which is dimension sensitive is
Lemma~3, because in $m$ dimensions the limiting equation analogous
to (22) reads
\be
H'' - (m-2) \coth(x) H' + 2 k H=0,
 \ee
so Lemma~3 is true iff  the solution $H(x)$ is oscillating at
infinity, that is, $4 k > (m-2)^2$. This imposes the condition
\be
 l>\frac{\sqrt{2}-1}{2} (m-2).
\ee Under this condition the proofs of Lemma~5 and Proposition~1
remain basically unchanged, thus we have
 \begin{theo}
For each odd $m \geq 3$ and $l$ satisfying the condition (43),
there exist a countable family of smooth solutions $f_n$ of
equation (40) satisfying the boundary conditions (10). The index
$n=0,1,2,...$ denotes the number of intersections of $f_n(\rho)$
with the line $f=\pi/2$ on $\rho\in[0,1)$.
\end{theo}
This theorem extends the recent result of Cazenave, Shatah, and
Tahvildar-Zadeh~\cite{csz} who proved existence of the ground
state solution $f_0$ in odd dimensions under the condition (43)
using variational methods.
\section*{Acknowledgments} I thank Robert Wald for discussions and
Arthur Wasserman for reading the manuscript and helpful remarks.
 This research was supported in part
by the KBN grant 2 P03B 010 16.

\end{document}